\documentclass[11pt]{article}
\def\IFUM{654}

\usepackage{amsfonts}
\usepackage{a4}
\usepackage{theorem}
\usepackage{amsmath}
\usepackage{amssymb}

\def\Red{}
\def\Black{}
\def\Blue{}

\newcommand{\myfrac}[2]{
	\begin{array}{l}
	{\!{#1}\!}\\\hline
 	{\!{#2}\!}
	\end{array}\,
}
\newcommand{\der}[2]{
	\myfrac{\delta #1}{\delta #2}
}

\newcommand{\st}[1]{\int d^4 x \, \left (
                        c      {\der{#1}{\varphi}} +
                        B      {\der{#1}{\bar c}} +
                        \theta_k {\der{#1}{\rho_k}} \right ) }
\def\G{\Gamma}
\def\GT{\tilde\G}

\newcommand\QAPX[1]{{\Delta_{#1}(x)\cdot\G}\Black}
\newcommand\QCAP[1]{{\Delta_{#1}}\Black}
\newcommand\hepth[1]{{\tt hep-th/#1}}
\newtheorem{prop}{Proposition}[section]
\newtheorem{lemma}{Lemma}[section]
\newtheorem{definition}{Definition}[section]
\newcommand{\props}[2]{    
        \begin{prop}{\label{prop:#1} #2}\end{prop}}    
\newcommand{\lemmas}[2]{    
        \begin{lemma}{\label{lem:#1} #2}\end{lemma}}    
\newcommand{\definitions}[1]{    
        \begin{definition}{#1}\end{definition}}    
\newcommand{\lem}[1]{lemma~{\bf\ref{lem:#1}}}    

\begin{document}

\title{
\begin{flushright}
{\bf \small IFUM \IFUM/FT} \\
\end{flushright}
\Red
An Approach to the Equivalence Theorem
by \\the Slavnov-Taylor Identities
\Black
}
\author{
 Ruggero Ferrari,
 Marco Picariello,
 Andrea Quadri\thanks{emails: Ruggero.Ferrari@mi.infn.it,
 Marco.Picariello@mi.infn.it, Andrea.Quadri@mi.infn.it}
\\
\small  Dipartimento di Fisica, Universit\`a di Milano,
	via Celoria 16, 20133 Milan, Italy \\
\small and  INFN, Sezione di Milano, Italy \\
}

\maketitle

%

\abstract{
\Blue
We discuss the Equivalence Theorem (ET) in the BRST formalism.
In particular the Slavnov-Taylor (ST) identities, derived at the
 formal level in the path-integral approach, are considered at the
 quantum level and are shown to be always anomaly free.
\par
Some discussion is devoted to the transformation of the fields;
 in fact the existence of a local inverse (at least as a formal power
 expansion) suggests a formulation of the ET, which allows a nilpotent
 BRST symmetry.
This strategy cannot be implemented at the quantum level if the
 inverse is non-local.
In this case we propose an alternative formulation of the ET, where,
 by using Faddeev-Popov fields, this difficulty is  circumvented.
In fact this approach allows the loop expansion both in the original
 and in the transformed theory.
In this case the algebraic formulation of the problem can be
 simplified by introducing some auxiliary fields.
The auxiliary fields can be eliminated by using the method of
 Batalin-Vilkovisky.
\par
We study the quantum deformation of the associated ST identity and
 show that a selected set of Green functions,
which in some cases can be identified with the
 physical observables of the model, does not depend on the choice of the
 transformation of the fields.
The computation of the cohomology for the classical linearized ST
 operator is performed by purely algebraic methods.
We do not rely on power-counting arguments.

In general the transformation of the fields yields a
 non-renormalizable theory.
When the equivalence is established between a renormalizable and a
 non-renormalizable theory, the ET provides a way to give a meaning to
 the last one by using the resulting ST identity.
In this case the Quantum Action Principle cannot be of any help in the
 discussion of the ET.
We assume and discuss the validity of a Quasi Classical Action
 Principle, which turns out to be sufficient for the present work.
\par
As an example we study the renormalizability and unitarity of massive
 QED in Proca's gauge by starting from a linear Lorentz-covariant
 gauge.
\Black
}
\vskip 0.5 truecm

Keywords: Equivalence Theorem, BRST
\newpage
%
\section{Introduction}

The equivalence theorem~(ET) has a long
history~\cite{Dyson:1948,Case:1949a,Case:1949b,Foldy:1951,Kallosh:1972ap,Kamefuchi:sb,Appelquist:tg},
 conventionally starting
from the alleged physical equivalence of the pseudoscalar and
pseudovector coupling in the pion-nucleon system.
Since then the significance of the statement has been enlarged and more
sophisticated techniques have been used to prove the theorem.
The ET is still a subject of some interesting
 researches~\cite{Gomis:1996jp,Stora,Blasi:1999ph,Tyutin:2000ht}.
The theorem follows from the fact that one can pass from a
 given theory to a second one by a change of the fields in the
 Lagrangian.
This is well-known, but to our knowledge it
has not been
 considered in full generality.
Moreover all known proofs of the ET are based on the validity of
the Quantum Action Principle~(QAP),
which is valid for power-counting renormalizable theories.
In the pioneer papers it is assumed either that there is 
 only one derivative for each field in the vertex (BPHZ
 framework)~\cite{Lam:1973qa,Bergere:1975tr} or that the Feynman
 propagators are standard ($(p^2-m^2)^{-1}$ for bosons and
$(\slash{\!\!\!p}-m)^{-1}$ for fermions)
and there are no derivative vertices (dimensional
 regularization)~\cite{Breitenlohner:1977hr}. In~\cite{Clark:1976ym}
 the proof of QAP is extended to the massless case for a
 power-counting renormalizable theory.
\par
Thus typically the proofs rely on the power-counting
 renormalizability in order to avoid
Lagrangian terms causing the theory to become non-renormalizable.
A generalization of these results is often made for non-renormalizable
 theories, i.e. it is assumed a  straightforward extension of the results
 of the QAP to non-renormalizable theories.

In perturbative QFT, the renormalization problems to be solved result
 into rather disturbing complexity.
It is a goal of this note to gather the cumulative experience of the
 past two decades to propose a step to simplify the problem by
 introducing the concept of Quasi Classical Action Principle (QCAP)
 (\cite{Stora2000} and appendix~{\bf\ref{app:QCAP}})
 and the linear Slavnov Taylor (ST)
 identities.
\par
One of the major advances in Quantum Field Theory (QFT)
 has been the introduction of Feynman's path-integral
 quantization.
Starting from a classical action $S[\Phi]$ (in our notations we use
 only Lorentz scalars, although the formalism can be applied to more
 general cases), Green functions are perturbatively constructed
 according to Gell-Mann and Low's expansion
 formula~\cite{Itzykson:1980rh}.
{\em Perturbative QFT} means here that the quantities of interest are
 computed order by order in $\hbar$ which is the loop expansion 
parameter, without any attempt to give a meaning to the complete sum.
In terms of Feynman's path-integral representation
 the formal expression of the perturbative
 expansion of the expectation value of the product of local operators
 ${\cal O}_i(\Phi(x_i))$ is
\begin{eqnarray}\label{eq:GellMann}
 \left\langle T \left(\prod_{i=1}^n{\cal O}_i(\Phi(x_i))\right)
 \right\rangle
 \propto
 \int [{\cal D}\Phi] \, \left(\prod_{i=1}^n{\cal O}_i(\Phi(x_i))\right)\,
 e^{i S[\Phi]} \, .
\end{eqnarray}
The properties of Feynman's path integral can be proven only at the
 formal level.
Thus the power of the formalism is to provide relations among the
 Green functions which often need a more sound formulation.
\\
It was early recognized by Becchi, Rouet, Stora and Tyutin (BRST) in
 the works~\cite{Becchi:1974xu,Becchi:1975nq,T}
 on BRST invariance of gauge theories and on the
 models with broken rigid symmetries~\cite{Becchi:1981jx}, that the
 use of the Quantum Action Principle
 leads to the possibility of a fully algebraic proof of
 the renormalizability of a theory characterized by a set of local or
 rigid invariances.
\\
Since then it has become a common paradigm in QFT to write Ward
 identities by using the path-integral and then use powerful methods
 of perturbation theory to prove them (as BPHZ renormalization scheme,
 BRST symmetry, etc ...).
\par
Let the classical action $S[\Phi]$ be a local functional of the field
 $\Phi$ (and its derivative) with non degenerate quadratic part.
Let $\{\G[\Phi,\beta]\}$ be the set of quantum extensions of
 $S[\Phi]$.
$\G[\Phi,\beta]$ is
 a formal power series in $\hbar$, the loop counting parameter,
 where the $\beta$ are introduced in order to generate the
Green functions in eq.(\ref{eq:GellMann}). Their connected part
is obtained by differentiating 
 the generating functional given by the Legendre
 transform of the quantum action $\G$ with respect to the field
 $\Phi$:
$$
 W[J,\beta]=\G[\Phi,\beta]+\int d^4x\ J(x)\Phi(x) \, .
$$

In renormalizable field theories the ambiguities in the definition of
 $\G$ are removed by power-counting and normalization condition.
Now suppose that we wish to parameterize the classical action in a
 different way, i.e. we wish to perform the following field
 redefinition:
\begin{eqnarray}\label{eq:redefinition}
 \Phi \rightarrow \Phi = \Phi(\varphi;\rho)
 &{\rm with}& \Phi(\varphi;0)=\varphi \, ;
\end{eqnarray}
 with $\rho = \{ \rho_i \}$ external classical fields parameterizing
 the transformation.
How is eq.~(\ref{eq:GellMann}) affected by this field redefinition? In
 particular, do the physical observables depend on the choice of the
 field parameterization?

If one regards the above transformation as a variable substitution in
 the path integral formulation of eq.~(\ref{eq:GellMann}) then the ET
 becomes
\begin{eqnarray}\label{eq:statSET}
 \int [{\cal D}\Phi] \,
  \left(
   \prod_{i=1}^n {\cal O}_i\left(\Phi(x_i)\right)
  \right)
  e^{i S[\Phi]}
& = & \int [{\cal D}\varphi] \,
  \left(
   \prod_{i=1}^n {\cal O}_i\left(\Phi(\varphi;\rho)(x_i)\right)
  \right)
  e^{i S[\Phi(\varphi;\rho)] + i D(\varphi;\rho)}\,,\nonumber \\
\end{eqnarray}
 where $D(\varphi;\rho)$ originates from the Jacobian of the field
 redefinition, and ${\cal O}$ is any local operator.
We denote this formulation of the ET as the Strong Equivalence Theorem
 (SET).
In order to give meaning to the r.h.s. of eq.~(\ref{eq:statSET}) a set
 of external sources will be necessary as for instance a term involving
 the new field $\varphi$ and the Faddeev-Popov ghosts which will be
 introduced in order to take into account the determinant.

It has been noticed in~\cite{Bergere:1975tr}
 that a simpler formulation of the ET can be
 given if one neglects the Jacobian of the transformation.
In this case the ET takes the form
\begin{eqnarray}\label{eq:statWET}
 \int [{\cal D}\Phi]\,
  \left(
   \prod_{i=1}^n {\cal O}_i\left(\Phi(x_i)\right)
  \right)
  e^{i S[\Phi]}
& = & \int [{\cal D}\varphi] \,
  \left(
   \prod_{i=1}^n {\cal O}_i\left(\Phi(\varphi;\rho)(x_i)\right)
  \right)
  e^{i S[\Phi(\varphi;\rho)]}\,,
\end{eqnarray}
 which we call the Weak Equivalence Theorem (WET).
This has been considered in different ways by previous
 authors~\cite{Tyutin:2000ht,Bergere:1975tr,Breitenlohner:1977hr}.
In some cases (see discussion below about $\rho$ expansion and loop
 expansion) it yields a set of local (counter)terms which can be
 reabsorbed by a redefinition of the action.
However some particularly interesting cases (those with non-local
 inverse transformation) cannot be dealt with in this way.
\\
The WET amounts to saying that $\Phi$ and $\varphi$ are physically
equivalent
 interpolating fields in the LSZ formalism.

The content of eq.~(\ref{eq:statWET}) can be discussed in the
 framework of BRST theory where the transformations are chosen in such
 a way that $\Phi(\varphi;\rho)$ is an
 invariant~\cite{Stora,Tyutin:2000ht}
\begin{eqnarray}\label{eq:sphizero}
 s \Phi(\varphi;\rho) = 0\, .
\end{eqnarray}
The transformation of $\varphi$ and $\rho_k$ should then be
\begin{eqnarray}\label{sPhi}
 s \varphi =
   - \left(  \der{\Phi}{\varphi} \right)^{-1}
              \der{\Phi}{\rho_k}\ \theta_k
\, , &\quad\quad&
 s \rho_k = \theta_k\,, \quad\quad s\theta_k=0\,.
\end{eqnarray}
This transformation is required to be formally local as a power series in
 $\rho_k$.

We now consider the Strong Equivalence Theorem (SET).
It states that the Green functions generated
 by the functional
\begin{eqnarray}\label{eq:ZstatSET}
 Z[J]
&\propto&
 \int [{\cal D}\Phi] \exp\left(
	i S[\Phi]
	+ i \int d^4x\,J\Phi \right)
\end{eqnarray}
 are related to those obtained from
\begin{eqnarray}\label{eq:tildeZ}
 \tilde Z[J,K;\rho]
&\propto&
 \int [{\cal D}\varphi] \exp\left(
	i S[\Phi(\varphi;\rho)]
	+ i D(\varphi;\rho)
	+ i \int d^4x\,J\Phi(\varphi;\rho)
        + i \int d^4x\,K\varphi \right)
\,,\nonumber\\
\end{eqnarray}
 where $D(\varphi;\rho)$ originates from the Jacobian of the field
 transformation.
Roughly speaking, eq.~(\ref{eq:statSET}) could be summarized by saying
 that the Green functions are insensitive to the fields
 reparameterization: once the classical action is fixed, the
 parameterization of the fields which we use in the path-integral is a
 matter of convenience.

As in the WET case, the content of eq.~(\ref{eq:statSET}) can be
 discussed in the framework of BRST theory.
However this problem has no straightforward solution.
We find more convenient to introduce a set of auxiliary fields, which
 allows a linearization of the ST identities.
We call it the {\it off-shell} formalism.
Eventually we shall provide also the BRST transformations relevant for
 the {\it on-shell} theory which yields the generating functional in
 eq.~(\ref{eq:tildeZ}).

One should remark that even if the l.h.s. of eq.~(\ref{eq:statSET}) refers
to a
 renormalizable theory, the action in the r.h.s. could be
 not-renormalizable and therefore the
 question will soon arise whether it is possible to constrain the
 renormalization ambiguities in such a way that eq.~(\ref{eq:statSET})
holds.
In this sense one is giving a procedure for treating a new class of
theories:
the non-renormalizable theories obtained from a renormalizable one after
 the reparameterization of eq.~(\ref{eq:redefinition})~\cite{Gomis:1996jp}.

The proof given in~\cite{Bergere:1975tr} for the Weak Equivalence
Theorem and in~\cite{Lam:1973qa} for the Strong Equivalence Theorem
 heavily relies on the properties of the BPHZ subtraction procedure.
It shows that if the $T$-products are defined according to the
BPHZ prescription with some suitably chosen subtraction indices,
 then eq.~(\ref{eq:statSET}) for the Strong Equivalence Theorem (and
 respectively eq.~(\ref{eq:statWET}) for the Weak Equivalence Theorem)
 holds true order by order in the perturbative expansion in $\rho$.
The issue of the validity of these results within different
 definitions of the $T$-products was not addressed.

Several algebraic aspects of the Equivalence Theorem in the BRST framework
  have been discussed in~\cite{Bastianelli:1990ey,Bastianelli:1991yk,Alfaro:1989xx,Alfaro:1989rx,Alfaro:1992cs,Alfaro:1993ua}.
Recently the Weak Equivalence Theorem has been investigated
 by using the BRST techniques in various ways,
 first in~\cite{Blasi:1999ph} and then in a different context
 in~\cite{Tyutin:2000ht}.

In Ref.~\cite{Tyutin:2000ht} the Equivalence Theorem
 is shown to hold true provided that both the transformation
 in eq.~(\ref{eq:redefinition}) and its inverse
 $\varphi = \varphi(\Phi;\rho)$ are local (at least perturbatively in
 the expansion in $\rho$).

In our work we prefer to perform a perturbative expansion in term of
 $\hbar$, which gives also the loop expansion.
In particular, the $\rho$ expansion is not equivalent to the loop
expansion in
 the class of transformations
\begin{eqnarray}\label{eq:redefinitionA}
 \Phi = A_{\rho} \varphi + \rho \Delta(\varphi)
\end{eqnarray}
 where $A_{\rho}$ is a $\rho$-dependent linear operator with
non-local
 inverse (e.g. $A_{\rho} = 1 + \rho_0 \square_x$) and
 $\Delta(\varphi)=\{\Delta_i(\varphi)\}$ accounts for
 the non linear part and is assumed to be a set of
 Lorentz-invariant monomials in $\varphi$ and its derivatives.
In this case the bilinear part of the action is substantially modified
 and thus the propagators are $\rho$-dependent in such a way that the
 loop expansion differs from the $\rho$ expansion.

This in turn extends the validity of the Equivalence Theorem
 to a much wider class of field redefinitions, including those
 in eq.~(\ref{eq:redefinitionA}).

Both in the strong and in the weak formulation, the classical invariance
under the BRST transformations is translated into ST identities for the
vertex functional.
Since ST identities are the ideal tools in order to study the quantum
extensions of these symmetries, it is crucial to prove that they are
anomaly free.
In the present paper we prove that indeed the ST identities can be
restored at every order in the loop expansion by using the techniques
of removing from the cohomology the terms depending on the ST doublets.

It is interesting to see how this can happen even in the case where
the theory is a gauge theory affected by an anomaly in the ST
identities associated to the gauge BRST invariance.
In this rather peculiar case however the field theory remains sick
since unitarity cannot be recovered~\cite{Picariello:2000xc}.

The paper is organized as follows.
In sect.~{\bf\ref{sec:WET}} we discuss the Weak Equivalence Theorem.
We derive the ST identities and we study
their quantum extension.
In sect.~{\bf\ref{sec:classical}} we discuss the Strong Equivalence
Theorem by using the path-integral approach.
In sect.~{\bf\ref{sec:off}} we consider the off-shell formalism and
 construct a suitable BRST symmetry under which the new action
 is invariant.
In sect.~{\bf\ref{sec:SToff}} the quantum deformation of the ST identities
 associated to the off-shell case are discussed.
In sect.~{\bf\ref{sec:STon}} we give an on-shell formulation of
 the Strong Equivalence Theorem and we study the quantum extension of
 the Slavnov-Taylor identities.
In sect.~{\bf\ref{sec:rho:indep}} we show that the $\rho$-independence of
a selected set of Green functions
(to be identified in some cases with the
 physical content of the theory)
is a consequence of the quantum ST identity.
In sect.~{\bf\ref{sec:example}} we give some applications of the
 Equivalence Theorem.
Conclusions are presented in sect.~{\bf\ref{sec:sez6}}.
In appendix~{\bf\ref{app:QCAP}} we discuss the Quasi Classical Action
Principle, and in appendix~{\bf\ref{app:doublets}} we study the
removal of the ST doublets from the cohomology.

\section{The Weak Equivalence Theorem}\label{sec:WET}

In this section we consider the Weak Equivalence Theorem as formulated
 in eq.~(\ref{eq:statWET}). At classical level we consider the
 vertex functional
\begin{eqnarray}\label{eq:G0}
 \G^{(0)}[\Phi,\beta] &=&
		S[\Phi]
	      + \int d^4x\, \beta_i {\cal O}_i(\Phi)
\, .
\end{eqnarray}
In the spirit of the WET, defined by eq.~(\ref{eq:statWET}),
 we perform a change of variable, of the kind given in
 eq.~(\ref{eq:redefinition}), inside the classical vertex
 functional~\cite{Lam:1973qa}.
We write the transformation of the field as
\begin{eqnarray} \label{eq:redefinitionDelta}
 \Phi = A_\rho\varphi +\sum_j \rho_j \Delta_j(\varphi) \,,
\end{eqnarray}
 where $A_{\rho}$ is a differential operator reducing to the identity
 for $\rho$=$0$, and $\Delta_j(\varphi)$ are monomials in $\varphi$ and
 its derivatives of order at least two in $\varphi$.

By performing the field redefinition of
 eq.~(\ref{eq:redefinitionDelta}) in $\G^{(0)}$ we obtain the new vertex
 functional $\GT^{(0)}$
\begin{eqnarray}\label{eq:G0WET}
 \GT^{(0)}[\varphi,\beta;\rho] &=&
		S[\Phi(\varphi;\rho)]
	      + \int d^4x\, \beta_i {\cal O}_i(\Phi(\varphi;\rho))
\, .
\end{eqnarray}
It should be stressed that the new action is not obtained by a change
 of variables in the path-integral.
Therefore one should not expect the generating functionals to
 coincide.
However there is a very intriguing way, based on the BRST formalism,
 to formulate the (weak) equivalence of the transformed theory to the
 original one~\cite{Stora,Tyutin:2000ht}.

The classical vertex functional $\GT^{(0)}$ depends on the
 field $\varphi$ only through the combination given by $\Phi(\varphi;\rho)$.
In order to formulate the WET in terms of BRST invariance
 we introduce a set of transformations with the requirements
 that $\Phi(\varphi;\rho)$ is left invariant.
This is what we should expect, since $\Phi(\varphi;\rho)$ is the field
 combination that appears in eq.~(\ref{eq:statWET}).
The BRST transformation of $\varphi$ is dictated by
 the requirement that $s \Phi(\varphi;\rho)$=$0$:
\begin{eqnarray}\label{eq:sPhi}
 0 = s\Phi(x) &=& \int d^4y
\left(\der{\Phi(x)}{\varphi(y)}\ s \varphi(y)
 +    \der{\Phi(x)}{\rho_k(y)}\ s\rho_k(y)
\right)\\
 &\Rightarrow&
s \varphi(x) =
 -\int d^4\, y \left (
 \myfrac{\delta \Phi}{\delta \varphi} \right )^{-1}\!\!\!\!\!\!\!(x,y)\,\,
 \left(\int d^4z\ \der{\Phi(y)}{\rho_k(z)}\  s\rho_k(z)\right)\,.
\end{eqnarray}
We introduce a new set of anticommuting external sources
 $\theta=\{\theta_k\}$ and we take $s \rho_k =\theta_k$,
 $s \theta_k =0$.
By construction  $\GT^{(0)}[\varphi,\beta;\rho]$ is invariant
 under the BRST transformations.
Therefore it obeys the following classical ST identity (STI)
\begin{eqnarray}\label{eq:classicalSTWET}
 \int d^4x\,\left(
 \der{\GT^{(0)}}{\rho_k(x)}\ \theta_k(x)
 +
 \der{\GT^{(0)}}{\varphi(x)}\ s\varphi(x)
 \right)
 & = & 0  \,
\end{eqnarray}
where (in compact notations)
\begin{eqnarray}\label{eq:BRSTWET}
 s \varphi =
  - \left(\der{\Phi}{\varphi} \right)^{-1}
	  \der{\Phi}{\rho_k}\ \theta_k
 \,,\quad
 s \rho_k=\theta_k
 \,,\quad
 s\theta_k=0\,.
\end{eqnarray}
A straightforward computation shows that $s^2 = 0$.
The Faddeev-Popov (FP) charge is assigned as follows:
\begin{eqnarray}\label{eq:FPWET}
 \mbox{FP}(\varphi) = \mbox{FP}(\rho_k) = 0 \, , \quad
 \mbox{FP}(\theta_k) = 1 \, .
\end{eqnarray}
The requirement that the vertex functional $\GT^{(0)}$ is FP neutral
 enforces the assignment of the FP charge to the remaining external
sources.

Notice that if the operator $\myfrac{\delta \Phi}{\delta \varphi}$ has no
 {\sl local} inverse (at least as a formal power series)
 the present formulation looses most of
 the nice properties related to BRST invariance.
We stress that, in order to extend eq.~(\ref{eq:classicalSTWET}) at
 quantum level, we need to couple $s \varphi$ to its antifield
 $\varphi^*$. This is possible only if $s \varphi$ is local as a
 formal power series.

\subsection{Quantum formulation of the WET}\label{sec:quantumWET}

In the sequel we shall discuss the quantum extension of
 eq.~(\ref{eq:classicalSTWET}). In particular we shall show
 that the classical STI in
 eq.~(\ref{eq:classicalSTWET}) can be fulfilled at the quantum level
 under the assumption that $\Phi(\varphi;\rho)$ is locally invertible
 as a formal power series in the fields and external sources and
 their derivatives.

Then it is straightforward to prove that the quantum
 extension of eq.~(\ref{eq:classicalSTWET}) gives the independence
 of $\rho$ of any Green function which depends only on
 $\Phi(\varphi;\rho)$, i.e. one obtains the statement of the
 WET provided by eq.~(\ref{eq:statWET}).

The WET allows to treat a set of theories, related by a change of fields,
 in which the Jacobian of the transformation does not play any r\^ole:
 this is a non-trivial feature of the locally invertible case, already
 put in evidence in the literature~\cite{Lam:1973qa,Bergere:1975tr}.
This feature cannot be properly understood by path-integral arguments.
This issue is here analyzed by using BRST techniques.

Once the BRST transformations of the fields $\varphi$ and
 the parameters $\rho$ and $\theta$ of eq.~(\ref{eq:BRSTWET}) are
 introduced, one
 needs to add to the vertex functional a term containing the antifield
 $\varphi^*$ coupled to $s \varphi$
\begin{eqnarray}\label{eq:G0WETstar}
\GT^{(0)}[\varphi,\theta,\varphi^*,\beta;\rho]
= S[\Phi(\varphi;\rho)]
       + \int d^4x \, \varphi^*\ s\varphi
       + \int d^4x \beta_i {\cal O}_i(\Phi(\varphi;\rho)) \, .
\end{eqnarray}
This is possible since $s \varphi$ is local
 as a formal power series in the fields and external sources
 and their derivatives.

The vertex functional $\GT^{(0)}$ satisfies the following
 Equivalence Theorem STI
\begin{eqnarray}\label{eq:STWET}
{\cal S}\left(\GT^{(0)}\right) \equiv \int d^4x \,
\left ( \myfrac{\delta\GT^{(0)}}{\delta\varphi^*}
        \myfrac{\delta\GT^{(0)}}{\delta\varphi} +
        \theta_k\myfrac{\delta\GT^{(0)}}{\delta\rho_k} \right ) =0\, .
\end{eqnarray}
Eqs.~(\ref{eq:G0WETstar}-\ref{eq:STWET}) are the starting point for
the quantization of the model.

\subsection{Quantum restoration of the STI}\label{sec:QuantumSTI}

The first task is then to show that the ET STI in eq.~(\ref{eq:STWET})
 can be preserved at the quantum level, i.e. we can define a quantum
 effective action $\GT$ by a suitable choice of non-invariant local
 counterterms, order by order in the loop expansion, such that
\begin{eqnarray}\label{eq:a:e12}
{\cal S}\left(\GT\right) =\int d^4x \,
\left ( \der{\GT}{\varphi^*}
        \der{\GT}{\varphi}+
\theta_k\der{\GT}{\rho_k} \right ) =0 \, .
\end{eqnarray}
The above STI has the same form as the well-known STI
 associated with the usual BRST symmetry, arising in ordinary gauge
 theories.
Therefore one expects to be able to characterize the $n$-th order
 ST breaking terms by algebraic constraints like the Wess-Zumino
 consistency condition, under the recursive assumption that the STI
 have been restored up to order $n$-$1$.

However the problem is much more intricate here, because almost
 surely the transformed theory, whose vertex functional is given by
 $\GT$, is not power-counting renormalizable.
Therefore the QAP cannot be applied here to guarantee that the
 $n$-th order ST breaking term is a local polynomial in the fields
 and the external sources and their derivatives.

We wish to comment on this point further.
For  power-counting renormalizable theories the QAP characterizes
 the possible breaking of the STI, to all orders in perturbation theory,
 as the insertion of a suitable local operator with bounded dimension.
Consequently, if such an insertion were zero up to order $n$-$1$,
 at the $n$-th order it must reduce to a local polynomial in the fields
 and the external sources and their derivatives with bounded dimension.
This follows from the locality of the insertion and the topological
 interpretation of the $\hbar$-expansion as a loop-wise expansion.

We stress that all what is needed to carry out the recursive analysis
 of the ST breaking terms by using algebraic methods is that part
 of the QAP, saying that at the first non-vanishing order
 in the loop expansion the ST breaking term is a local polynomial
 in the fields and the external sources and their derivatives.
Again, we wish to emphasize that indeed the QAP says much more,
 since it characterizes the ST breaking terms to all orders in perturbation
 theory as the insertion of a suitable local operator.
This is true independently of the normalization conditions
 and of the non-invariant finite action-like counterterms chosen.
It might eventually happen that a careful order by order choice of the
 latter allows to set such a local insertion equal to zero.
In this case we speak of a non-anomalous theory.

In the case of a non power-counting renormalizable theory, an all-order
 extension of the QAP could hardly be proven. This is because several
 restrictive assumptions on the form of the propagators and on the
 interaction vertices are required in all known proofs of the
 QAP.
None of them can be imposed with enough generality to non power-counting
 renormalizable theories.

The key observation~\cite{Stora2000,Piguet:1995er} is that the full
 power of the QAP is never used in discussing the recursive restoration
 of the STI.
One only needs the locality of the breaking terms at the first
 non-vanishing order in the loop expansion.
It has then been proposed~\cite{Stora2000} that a suitable extension of this
 statement of the QAP might be the good property to look for
 when dealing with non power-counting renormalizable theories.
Such an extension has been called by Stora the Quasi-Classical Action
 Principle.
It states that
\begin{quotation}
\noindent
{\em
In the loop-wise perturbative expansion the first non-vanishing order
 of ST-like identities is a classical local integrated formal power
 series in the fields and external sources and their derivatives.}
\end{quotation}
Let us comment on this proposal. Although its
 plausibility, no satisfactory proof is available
 for the QCAP.
Thus from now on we will assume that it is fulfilled by the regularization
 used to construct the vertex functional $\G$.
Moreover, we point out that we do not require that
 the breaking term is polynomial:
 loosing the power-counting entails that no bounds on the dimension
 can in general be expected.

Since we assume the QCAP, we can now use the power of cohomological
 algebra to constrain the possible ET ST breaking terms and show
 that the ET STI can always be restored by a suitable order by order
 choice of non-invariant counterterms.

The proof is a recursive one. Assume that the ET STI in eq.~(\ref{eq:a:e12})
 has been fulfilled up to order $n$-$1$:
\begin{eqnarray} \label{eq:a:e5}
 {\cal S}\left(\GT\right)^{(j)}=0 \, , ~~~~ j=0,\dots,n-1 \, .
\end{eqnarray}
At the next order the ET STI might be broken by the functional
 $\Delta^{(n)}$ defined by
\begin{eqnarray}\label{eq:a:e6}
 \Delta^{(n)} \equiv {\cal S}\left(\GT\right)^{(n)} \, .
\end{eqnarray}
Since we are assuming that the QCAP holds, we know that
 $\Delta^{(n)}$ is a local formal power series in the fields
 and external sources and their derivatives and has FP-charge $+1$.
No bounds on the dimensions can however be given, since
 in general we are dealing with a theory which is non-renormalizable
 by power-counting.
This in turn rules out the possibility to constrain
 $\Delta^{(n)}$ by means of power-counting arguments.

We then must resort to purely algebraic arguments
 in order to show that $\Delta^{(n)}$ can indeed be removed
 by a suitable choice of $n$-th order local non-invariant counterterms.
This is a viable way to characterize $\Delta^{(n)}$ because of
 the following Wess-Zumino consistency condition obeyed by $\Delta^{(n)}$:
\begin{eqnarray} \label{eq:a:e7}
 {\cal S}_0\left(\Delta^{(n)}\right)=0 \, ,
\end{eqnarray}
 where ${\cal S}_0$ denotes the classical linearized ST operator
\begin{eqnarray}\label{eq:a:e8}
 {\cal S}_0 = \int d^4x \,
 \left  ( \der{\GT^{(0)}}{\varphi^*}\der{}{\varphi}
         +\der{\GT^{(0)}}{\varphi}  \der{}{\varphi^*}
       +\theta_k \der{}{\rho_k}
 \right )  \, .
\end{eqnarray}
It can be easily proven that ${\cal S}_0$ is nilpotent,
 as a consequence of the classical STI in eq.~(\ref{eq:STWET}).
Thus a cohomological analysis of the functional $\Delta^{(n)}$,
 starting from eq.~(\ref{eq:a:e7}), is at hand.

We first notice that the sources $(\rho_k,\theta_k)$ form a set
 of doublets under ${\cal S}_0$.
In a situation where the counting operator
\begin{eqnarray}\label{eq:a:e10}
 {\cal N} = \sum_k \int d^4x \, \left
(
 \rho_k {\der{}{\rho_k}} +
 \theta_k {\der{}{\theta_k}}
 \right )
\end{eqnarray}
 does commute with ${\cal S}_0$, the doublets can be removed from
 the cohomology by standard techniques~\cite{Zumino:1983ew}.
Here however the doublets are coupled since
 $[{\cal N},{\cal S}_0]\neq 0$.
Under some conditions the coupled doublets can be also
 removed~\cite{doppietti}.
Here in particular 
$\Delta^{(n)}$ in eq.~(\ref{eq:a:e6}) is ${\cal S}_0$-invariant by
the Wess-Zumino consistency condition in eq.~(\ref{eq:a:e7}).
Moreover since there are neither fields nor external sources
with ${\rm FP}$-charge $+1$ apart from $\theta_k$, $\Delta^{(n)}$
must vanish at $\theta_k=0$. This yields
\begin{eqnarray}
\left . {\cal S}_0\left(\Delta^{(n)}\right |_{\theta_k=\rho_k=0}\right) = 0 \, .
\label{eq:z:e1}
\end{eqnarray}
In~\cite{doppietti} it is shown that the above equation,
 combined with eq.~(\ref{eq:a:e7}),
 is a sufficient condition
to conclude that the functional
$\Delta^{(n)} - \left . \Delta^{(n)} \right |_{\theta_k = \rho_k = 0}$
is cohomologically trivial, i.e.
there exists a local formal power series $\Xi^{(n)}$ such that
\begin{eqnarray}
\Delta^{(n)} -
\left . \Delta^{(n)} \right |_{\theta_k = \rho_k = 0}
= {\cal S}_0 \left(\Xi^{(n)}\right) \, .
\label{eq:z:e2}
\end{eqnarray}
Moreover, since $\left . \Delta^{(n)} \right |_{\theta_k = \rho_k = 0}=0$,
we get from eq.~(\ref{eq:z:e2}) that the whole $\Delta^{(n)}$ is
${\cal S}_0$-invariant.

By adding to the $n$-th order effective action $\GT^{(n)}$ the $n$-th order
non-invariant counterterm $-\Xi^{(n)}$ we conclude from eq.~(\ref{eq:z:e2})
that the STI can be restored at the $n$-th order, i.e.
\begin{eqnarray}
{\cal S}\left(\GT\right)^{(n)} = 0 \, .
\label{eq:z:e3}
\end{eqnarray}
This ends the proof that the ET STI in eq.~(\ref{eq:a:e12}) can be fulfilled
to all orders
in the loop expansion:
\begin{eqnarray}
{\cal S}\left(\GT\right) = \int d^4x \,
\left(
	\der{\GT}{\varphi^*}\der{\GT}{\varphi} 
	+ \theta_k \der{\GT}{\rho_k}
\right) \, .
\label{eq:z:e4}
\end{eqnarray}
%

\section{The Strong Equivalence Theorem: a path-integral formulation}
\label{sec:classical}
The main feature of the Strong Equivalence Theorem is that it is obtained
 by performing a change of variables in the path integral, as already
 pointed out in eqs.~(\ref{eq:ZstatSET}) and~(\ref{eq:tildeZ}) .

As in the WET case, the content of eq.~(\ref{eq:statSET}) is here
 discussed in the framework of BRST theory.
However this problem has not straightforward solution.

We find more convenient to introduce a set of auxiliary fields, which
 allows a linearization of the ST identities.
We call it the {\it off-shell} formalism.
We will prove that the related STI can be fulfilled at the quantum level.
From the quantum STI, we will deduce the statement of the SET in the
 off-shell formalism.

We shall also provide the STI relevant for
 the {\it on-shell} formulation of the equivalence between the
 theories given by the generating functionals in
 eqs.~(\ref{eq:ZstatSET}) and~(\ref{eq:tildeZ}).
The corresponding quantum STI will allow us to prove the validity of
 the statement of the SET defined by eq.~(\ref{eq:statSET}).

In this section we study the Strong Equivalence Theorem
 from the point of view of the path-integral formalism.
In this framework the field redefinition in
 eq.~(\ref{eq:redefinition})
 can be naturally interpreted as a change of variables.
This turns out to be very useful in order to derive at the formal
 level the ST identities controlling the renormalization of the
 transformed theory.

In the path-integral formalism, one starts from the classical
 action $S[\Phi]$ and defines a generating functional
\begin{eqnarray}\label{eq:Zbeta}
 Z[J,\beta] = \int [{\cal D}\Phi] \,
 \exp \left (i S[\Phi] + i \int d^4x \, J \Phi
 + i \int d^4x \, \beta_i {\cal O}_i(\Phi)   \right ) \, .
\end{eqnarray}
In the above equation we have introduced the external
 sources $\beta_i$, coupled to the local operators
 ${\cal O}_i(\Phi)$.

We now perform the field redefinition
\begin{eqnarray} \label{eq:redefinitionDeltaSET}
 \Phi = A_{\rho} \varphi + \sum_j \rho_j \Delta_j(\varphi) \,.
\end{eqnarray}
Performing the change of variables of eq.~(\ref{eq:redefinitionDeltaSET})
 in eq.~(\ref{eq:Zbeta}) one gets
{
\begin{eqnarray}\label{et::re5}
\!\tilde Z[J,\beta] & = & \int
	[{\cal D} \varphi]
	\, {\rm det} \, \left ( \der{\Phi}{\varphi} \right )
	\exp \left ( i S[\Phi(\varphi;\rho)]
		   + i \int d^4x \, J \Phi(\varphi;\rho)
		   + i \int d^4 x \, \beta_i {\cal O}_i(\Phi(\varphi;\rho))
	     \right )
 \nonumber \\
& = &  \lim_{K,\eta,\bar\eta  \rightarrow 0}
 \tilde Z_{on}[J,\beta, K, \eta, \bar\eta ;\rho]
\end{eqnarray}
}
where we have introduced:
\begin{eqnarray}\label{eq:TildeZOn}
\tilde Z_{on}[J,\beta, K , \eta, \bar\eta;\rho] &=&
\int [{\cal D} \varphi] [{\cal D} \bar c][{\cal D} c]
	\exp \left ( i S[\Phi(\varphi;\rho)]
           + i \int d^4x \, \bar c
		\left (
			\myfrac{\delta \Phi}{\delta \varphi} c
		      + \myfrac{\delta \Phi}{\delta \rho_k} \theta_k
		\right )
 \right.  \nonumber
\\&& \left. \quad
	   + i \int d^4x \, J \Phi(\varphi;\rho)
           + i \int d^4x \, \beta_i {\cal O}_i(\chi)
 \right.  \nonumber
\\&& \left. \quad
           + i \int d^4x \,  K \varphi
	   + i \int d^4x\, \eta\bar c
	   + i \int d^4x\, \bar \eta c
\right ) \, .
\end{eqnarray}
To deal with the Jacobian of the transformation, we
 use the standard trick of introducing the Faddeev-Popov ghost
 and antighost $c,\bar c$ to exponentiate the determinant.
For later convenience we have also added to the action the term
 $$i\int d^4x\, \bar c \myfrac{\delta \Phi}{\delta \rho_k} \theta_k\,.$$
The external sources $\theta_k$ have opposite statistics with respect
 to $\rho_k$ and will be interpreted as before as the BRST partners of
 $\rho_k$.
In eq.~(\ref{eq:TildeZOn}) we need the sources $ K $, $\eta$
 and $\bar\eta$ coupled to the quantum fields $\varphi$, $\bar c$ and $c$,
 in order to define the perturbative expansion of the theory.

The external source $J$ is now associated to the
 {\em composite} (from the point of view of $\varphi$) operator
 $\Phi(\varphi;\rho)$. It is on the same footing as
 the  external sources $\beta$.
The limit $ K,\eta,\bar\eta  \rightarrow 0$
 means that in the end we care about amplitudes
 with $\Phi$ or ${\cal O}(\Phi)$ external legs.

We will prove in sect.~{\bf\ref{sec:STon}}, by using the method of
 Batalin-Vilkovisky, that the construction of the Green functions can be
 equivalently performed via the generating functional $Z$ or $\tilde
 Z_{on}$.
This is the on-shell version of the SET.

A less involved formulation, from an algebraic point of view,
 is obtained by introducing a couple of auxiliary fields
 $B$ and $\chi$.
That is, we rewrite $Z[J,\beta]$ as
{
\begin{eqnarray}\label{et::re5bis}
\tilde Z[J,\beta] & = &
\int [{\cal D} \varphi] [{\cal D} \chi] \delta(\chi-\Phi(\varphi;\rho))
\, {\rm det} \, \left ( \der{\Phi}{\varphi} \right )
\exp \left ( i S[\chi]
           + i \int d^4x \, J \chi
           + i \int d^4 x \, \beta_i {\cal O}_i(\chi)
\right)
 \nonumber \\
& = &  \lim_{ K ,J_B,\eta,\bar\eta \rightarrow 0}
 \tilde Z_{off}[J,\beta, K ,J_B,\eta,\bar\eta;\rho]
\end{eqnarray}
}
 where now
{
\begin{eqnarray}\label{eq:TildeZOff}
\!\tilde Z_{off}[J,\beta, K ,J_B,\eta,\bar\eta;\rho] &=&
\int [{\cal D} \varphi] [{\cal D} \chi] [{\cal D} B]
     [{\cal D} \bar c][{\cal D} c]
\exp \left(  i S[\chi]
           + i \int d^4x B(\chi - \Phi(\varphi;\rho))
 \right.  \nonumber \\
&& \left. \quad
           + i \int d^4x \, \bar c
		\left (
			\myfrac{\delta \Phi}{\delta \varphi} c
		      + \myfrac{\delta \Phi}{\delta \rho_k} \theta_k
		\right )
           + i \int d^4x \, J\chi
	   + i \int d^4x\, \eta\bar c
 \right.  \nonumber \\
&& \left. \quad
	   + i \int d^4x\, \bar \eta c
           + i \int d^4x \,  K \varphi + i \int d^4x J_B B
           + i \int d^4x \, \beta_i {\cal O}_i(\chi)
\right ) \, .
\nonumber \\
\end{eqnarray}
We speak in this case of the off-shell formulation of the Equivalence
 Theorem.
In eq.~(\ref{eq:TildeZOff}) $ K $ and $J_B$  are respectively
 the sources coupled to $\varphi$ and $B$.
Notice that $J$ is now coupled to $\chi$.
As much as in the on-shell case, there exists a ST
 identity expressing the $\rho$-independence of
 the Green functions involving solely $\chi$ and ${\cal O}_i(\chi)$.
Moreover, the ST identity is linear in the off-shell case.
Despite of this advantage, the calculations in the off-shell scheme
 are clumsy due to the high number of quantized fields involved.
We will illustrate the derivation of the ST identities
 in the off-shell case in sect.~{\bf\ref{sec:off}}.

The on-shell and off-shell cases can be related by eliminating the
 auxiliary fields. This will be discussed in
 sect.~{\bf\ref{sec:SToff}}.

\section{The off-shell fomulation of the SET}\label{sec:off}

\subsection{The classical Slavnov-Taylor identities}

We discuss here the construction of the BRST transformations and
 related ST identities in the off-shell case.
For that purpose we need to work on the Legendre transform
 $\GT_{off}$ of
 the connected generating functional
 $\tilde W_{off}$, where we
 have defined
\begin{eqnarray}
\tilde Z_{off}= \exp
\left ( \frac{i}{\hbar} \tilde W_{off}
 \right ) \, .
\end{eqnarray}
From eq.~(\ref{eq:TildeZOff}) we deduce the classical vertex functional
 $\GT^{(0)}_{off}$
\begin{eqnarray}\label{eq:offshellaction}
\! \GT^{(0)}_{off} & \equiv & S[\chi]
       + \int d^4x B(\chi - \Phi(\varphi;\rho))
       + \int d^4x \bar c \left (
              \myfrac{\delta \Phi}{\delta \varphi} c
            + \myfrac{\delta \Phi}{\delta \rho_k} \theta_k
                   \right )
       +  \int d^4 x \, \beta_i {\cal O}_i(\chi) \, .
\end{eqnarray}
The BRST differential can be easily written by looking at the
 classical vertex functional $\GT^{(0)}_{off}$.
Let us define
\begin{eqnarray}
&&
s \varphi = c, ~~~~ s \bar c = B, ~~~~ s \rho_k = \theta_k, ~~~~
s \chi = 0, \nonumber \\
&&
s c =0, ~~~~ sB=0, ~~~~ s\theta_k=0 \, .
\label{et::cheq8}
\end{eqnarray}
We assign the corresponding FP-charge by setting
\begin{eqnarray}\label{et::cheqFP}
{\rm FP}(\varphi)= {\rm FP}(B) = {\rm FP}(\rho_k) = {\rm FP}(\chi) = 0 \, ,
~~
{\rm FP}(c)={\rm FP}(\theta_k)=1 \, , ~~~~ {\rm FP}(\bar c)=-1 \, .
\end{eqnarray}
The FP-charge of the external sources can be read off by imposing
the requirement that
the classical action $\GT^{(0)}_{off}$ is FP-neutral.
The vertex functional $\GT^{(0)}_{off}$ in
eq.~(\ref{eq:offshellaction}) can then be cast in the form
\begin{eqnarray}
\GT^{(0)}_{off}  =
S[\chi] + s \int d^4 x \bar c \left(\chi - \Phi(\varphi;\rho)\right)
+  \int d^4 x \, \beta_i {\cal O}_i(\chi)
\, .
\label{et::cheq9}
\end{eqnarray}
With this procedure one has automatically $s S[\chi]=0$ and
 $s {\cal O}_i(\chi)=0$.
We understand that the operators ${\cal O}_i(\Phi)$ now
become the operators ${\cal O}_i(\chi)$.
$\GT^{(0)}_{off}$ is $s$-invariant.
A similar approach has been adopted in~\cite{Alfaro:1989rx} where the issue
  of the cohomology is not dealt with.

Notice that in this formulation the whole $\Phi$ dependence appears
 through the cohomologically trivial term
\begin{eqnarray*}
 \int d^4x \, B(\chi - \Phi(\varphi;\rho)) +
\int d^4x \, \bar c \left(
{\der{\Phi}{\varphi}} c
+ \der{\Phi}{\rho_k} \theta_k \right )=
s \int d^4 x\,  \bar c \left(\chi - \Phi(\varphi;\rho)\right)
\end{eqnarray*}

The change of variable in the path integral has given rise
to an ET ST identity of the transformed action:
\begin{eqnarray}
{\cal S}\left(\GT^{(0)}_{off}\right) = \st{\GT^{(0)}_{off}} = 0 \, .
\label{et::cheq11_0}
\end{eqnarray}
This ET STI is linear, hence it does not require the introduction of
 any antifield.
Moreover, it can be seen from eq.~(\ref{et::cheq11_0}) that all fields
 and external sources transforming under ${\cal S}$ form a set of
 (decoupled) doublets (\cite{doppietti} and
 appendix~{\bf\ref{app:doublets}}), given by $(\varphi,c)$, $(\bar c,
 B)$ and $(\rho_k,\theta_k)$.
Therefore the study of the cohomology of ${\cal S}$ is particularly
 simple.
This is a significant advantage allowing to  treat 
also those situations where $\Phi =
 \Phi(\varphi;\rho)$ does not have a local inverse, by using 
algebraic techniques.

\section{Quantization of the ET STI in the off-shell formalism}
\label{sec:SToff}

We now study the quantization of the ET STI
 in the off-shell case.
Things are much simpler here due to the fact that all fields and
 external sources entering in the ET STI form sets of
 ordinary~\footnote{decoupled, according to the terminology
 of~\cite{doppietti}} doublets.

The proof that the classical ET STI in eq.~(\ref{et::cheq11_0})
 can be extended to all orders in the loop expansion is as follows.
We assume that the ET STI have been restored up to order $n$-$1$, i.e.
\begin{eqnarray*}
 {\cal S}(\GT_{off})^{(j)}&=&\st{\GT_{off}^{(j)}} = 0 \, , \quad
 j=0,1,\dots,n-1 \, .
\end{eqnarray*}
Thanks to the QCAP at the $n$-th order the possible ST breaking terms
\begin{eqnarray*}
 \Delta^{(n)} \equiv {\cal S}(\GT_{off})^{(n)} = {\cal S}(\GT_{off}^{(n)})
 &=&\st{\GT_{off}^{(n)}} \,
\end{eqnarray*}
 are local integrated formal power series in the fields, the external
 sources and their derivatives.
Notice that ${\cal S}$ is linear in the quantum fields.
Therefore no antifields are needed in the off-shell formalism.
This in turn yields that $\Delta^{(n)}$ only depends on the $n$-th
 order vertex functional $\GT_{off}^{(n)}$ and does not receive contributions
 from lower order terms, unlike in the on-shell case.

The Wess-Zumino consistency condition reads:
\begin{eqnarray}\label{STdelta}
 {\cal S}{\Delta^{(n)}}&=&\st{\Delta^{(n)}} = 0 \, .
\end{eqnarray}

We observe that the counting operator for the ${\cal S}$-doublets
 $(\varphi,c)$, $(\bar c, B)$ and $(\rho_k,\theta_k)$ commutes with
 ${\cal S}$.
Therefore we can introduce a homotopy operator
 ${\cal K}$~\cite{Zumino:1983ew} trivializing the cohomology of
 ${\cal S}$.
${\cal K}$ is defined as
\begin{eqnarray}
 {\cal K} = \int_0^1 dt \int d^4x \, \left (
 \varphi \lambda_t \der{}{c} + \bar c \lambda_t \der{}{B}
 +\rho_k \lambda_t \der{}{\theta_k} \right )
 \label{eq:z:10}
\end{eqnarray}
 where $\lambda_t$ acts as follows on an arbitrary functional $X$
 depending on $\varphi,c,\bar c, B, \rho_k,\theta_k$ and on other
 variables~$\zeta$:
\begin{eqnarray}
 \lambda_t X[\varphi,c,\bar c, B, \rho_k,\theta_k,\zeta]
 = X[t\varphi,t c, t \bar c, t B, t \rho_k, t \theta_k,\zeta] \, .
 \label{eq:z:11}
\end{eqnarray}
By explicit computation one verifies that
\begin{eqnarray}
 \{ {\cal K}, {\cal S} \} X = X
 \label{eq:z:12}
\end{eqnarray}
 where in the above equation
 $X[\varphi,c,\bar c, B, \rho_k,\theta_k,\zeta]$ is assumed
 to vanish at $c$=$\theta_k$=$0$ (as it happens for $\Delta^{(n)}$, since
 it has FP-charge $+1$ and the only fields and external sources
 with FP-charge $+1$ are $c$ and $\theta_k$).
We apply eq.~(\ref{eq:z:12}) to $\Delta^{(n)}$ and use the Wess-Zumino
 consistency condition in eq.~(\ref{STdelta}). We arrive at
\begin{eqnarray}
 \Delta^{(n)} = \{ {\cal K}, {\cal S} \} \Delta^{(n)} =
 {\cal S}{\cal K}\Delta^{(n)}\, .
 \label{eq:z:13}
\end{eqnarray}
Therefore $-\Xi^{(n)}$=${\cal K}\Delta^{(n)}$ are the counterterms to
 be added to the $n$-th order quantum effective action
 $\GT_{off}^{(n)}$ to guarantee that the STI are fulfilled at the
 $n$-th order in the loop expansion.
Notice that by the QCAP $\Xi^{(n)}$ is local in the sense of
 integrated formal power series.

We have proven that the ET STI can be fulfilled in the off-shell formalism,
i.e. it is possible to define to all orders in the loop expansion
a quantum effective action $\GT_{off}$ such that
\begin{eqnarray}
{\cal S}(\GT_{off}) = \st{\GT_{off}} = 0 \, .
\label{eq:z:14}
\end{eqnarray}
The above equation ensures that the Green functions of
local (in the sense of formal power series) BRST invariant operators
are $\rho$-independent.
The proof is reported in sect.~{\bf\ref{sec:rho:indep}}.

Notice that every counter term of the form ${\cal M}$=${\cal M}(\chi)$ is
such that
$$
{\cal S}({\cal M})=\st{{\cal M}}=0\,;
$$
that is it can be safely added to the $n$-th order ST invariant
normalization
 conditions without altering the ST identity.
Thus the physical conditions imposed on the theory for $\rho$=$0$ can
be preserved in the theory with $\rho \neq 0$.

\section{The SET in the on-shell formalism}
\label{sec:STon}

The Strong Equivalence Theorem can be formulated in the framework of
 BRST invariance also without introducing auxilary fields (on-shell
 formalism).
However one needs to use the formalism of Batalin-Vilkovisky to full
 extent.
The starting point is the following on-shell classical action
(for convenience we leave out the $\tilde{ }$ and the subscript $_{on}$
 from $\G$):
\begin{eqnarray}
\G_0^{(0)} = S[\Phi(\varphi;\rho),\beta] + \int d^4x \, \bar c s
 \Phi(\varphi;\rho) \, ,
\label{e1}
\end{eqnarray}
where the superscript $^{(0)}$ denotes the zero-th order in the loop
expansion (the classical approximation), and the subscript $_0$ refers
to the order in the antighost $\bar c^*$
 (conjugated to the antighost
field $\bar c$) to be shortly introduced.
The full on-shell classical action $\G^{(0)}$ will
eventually include terms of all orders in $\bar c^*$.

$S$ depends on the field $\Phi$ and on a set of external sources
$\beta$ coupled to local composite operators ${\cal O}(\Phi)$, functions of
$\Phi$ only.

The BRST differential $s$ is defined as
\begin{eqnarray}\label{e4}
&& s \varphi = c \, , ~~~~ s c = 0 \, , ~~~~
s \rho = \theta\, , ~~~~ s \theta = 0 \, .
\end{eqnarray}
That is, $(\varphi,c)$ and $(\rho,\theta)$ enter as doublets under
$s$.
In order to achieve the BRST invariance of
$\G_0^{(0)}$ we need to define the BRST variation of $\bar c$ as
follows:
\begin{eqnarray}
s \bar c = - \der{S}{\Phi} \, .
\label{e2}
\end{eqnarray}
Since $s^2\Phi(\varphi;\rho)$=$0$ the above equation guarantees that
$\G_0^{(0)}$ is BRST invariant.

We remark that the BRST differential $s$ is nilpotent only on-shell,
since
\begin{eqnarray}
s^2 \bar c = - \der{^2S}{\Phi\delta\Phi}
\  s \Phi =  - \der{^2S}{\Phi\delta\Phi}
        \  \frac{\delta \G_{0}^{(0)}}{\delta \bar c}
 \, .
\label{e2_new}
\end{eqnarray}
As is well known, in order to define the
composite operator $s\bar c$ at the quantum level we need to couple it
 in the classical action to the corresponding antifield $\bar c^*$.
We denote by $\Sigma_1$ those terms of the classical action
of order one in $\bar c^*$:
\begin{eqnarray}
\Sigma_1 = - \int d^4 x \, \bar c^*
	\der{S}{\Phi} \, .
\label{e3}
\end{eqnarray}
The new classical action is
\begin{eqnarray}
\G_1^{(0)} = \G_0^{(0)} + \Sigma_1 \, .
\label{e3bis}
\end{eqnarray}
$\G_0^{(0)}$ is BRST invariant, however $\G_1^{(0)}$ is not:
\begin{eqnarray}
s \G_1^{(0)} & = &  - \int d^4x \, \bar c^*
 \der{^2S}{\Phi\delta\Phi} 
 s \Phi \nonumber \\
       & = &  - \int d^4x \, \bar c^*
 \der{^2S}{\Phi\delta\Phi}
                  \frac{\delta \G_1^{(0)}}{\delta \bar c}  \, .
\label{e5}
\end{eqnarray}
The breaking term in the r.h.s. of the above equation is linear in
$\bar c^*$.
The BRST invariance of the classical action is only obtained in the on-shell
limit
$\frac{\delta \G_0^{(0)}}{\delta \bar c} $=
$\frac{\delta \G_1^{(0)}}{\delta \bar c} $=$0$.

This difficulty can be circumvented by making full use of the
 Batalin-Vilkovisky formalism.
 First we define the
 ST operator as
\begin{eqnarray}\label{e6}
{\cal S} (X) & = &
\delta X
 + (X,X)
\end{eqnarray}
where the parenthesis is given by
\begin{eqnarray}\label{eq:On7}
(X,Y) = \int d^4x \, \myfrac{\delta X}{\delta \bar c^*}
		     \myfrac{\delta Y}{\delta \bar c}
\end{eqnarray}
and the short-hand $\delta$ is for the linear part of the ST operator
\begin{eqnarray}
\delta X =  \int d^4x \,
\left ( c \myfrac{\delta X}{\delta\varphi} + \theta \myfrac{\delta
X}{\delta\rho}
\right ) \, .
\label{e8}
\end{eqnarray}

Then, in order to implement at the classical level the ST identities
 without going on-shell with $\bar c$ (i.e. without imposing
 $\frac{\delta \G^{(0)}}{\delta \bar c} = s \Phi =0$)
 we add to $\G^{(0)}_0$ monomials of higher degree in $\bar c^*$.

Thus we will construct the full on-shell
 classical action $\G^{(0)}$ by providing all
 of its $\bar c^*$-dependent components  $\Sigma_j$. $\Sigma_j$ is
 assumed to be  of order $j$ in $\bar c^*$.
Moreover the $\Sigma_j$ fulfill for $j>0$
\begin{eqnarray}\label{ghost2}
\myfrac{\delta \Sigma_j}{\delta \bar c} = 0
\end{eqnarray}
so that
the full on-shell classical action
satisfies the ghost equation
\begin{eqnarray} \label{ghost}
 \frac{\delta \G^{(0)}}{\delta \bar c} =
\frac{\delta \G_0^{(0)}}{\delta \bar c}=
 s \Phi \, .
\end{eqnarray}
The functionals $\Sigma_j$ are constructed as follows.

Eq.(\ref{e5}) is translated into
\begin{eqnarray}\label{e7bis}
{\cal S}(\G_1^{(0)}) =  - \int d^4x \, \bar c^*
\der{^2S}{\Phi\delta\Phi}
                  \frac{\delta \G_1^{(0)}}{\delta \bar c} \, .
\end{eqnarray}
We begin by adding the term $\Sigma_2$ quadratic in $\bar c^*$:
\begin{eqnarray}
\Sigma_2 = \int d^4x \, \frac{1}{2} \left ( \bar c^* \right )^2
 \der{^2S}{\Phi\delta\Phi} 
\label{eq:On9}
\end{eqnarray}
and define
\begin{eqnarray}
\G_2^{(0)} = \Sigma_0 + \Sigma_1 + \Sigma_2 \, .
\label{eq:On10}
\end{eqnarray}
It turns out that ${\cal S}(\G_2^{(0)})$ is quadratic in $\bar c^*$:
\begin{eqnarray}
{\cal S}(\G_2^{(0)}) & = & {\cal S}(\G_1^{(0)}) 
               + \delta \Sigma_2  + (\G_1^{(0)}, \Sigma_2)
               + (\Sigma_2, \G_1^{(0)}) 
               + (\Sigma_2,\Sigma_2) \nonumber \\
	       & = & - \int d^4x \, \bar c^*
                  \der{^2S}{\Phi\delta\Phi}
                  \frac{\delta \G_2^{(0)}}{\delta \bar c} 
               + s \Sigma_2 
               + \int d^4x \, \bar c^*
                  \der{^2S}{\Phi\delta\Phi}
                  \frac{\delta \G_2^{(0)}}{\delta \bar c}
\nonumber \\
	       & = & s \Sigma_2 = \int d^4x \,
			 \frac{1}{2} (\bar c^*)^2
                  \der{^3S}{\Phi\delta\Phi\delta\Phi} 
 s \Phi
		     \nonumber \\
	       & = & \int d^4x \, \frac{1}{2}
		 (\bar c^*)^2
                  \der{^3S}{\Phi\delta\Phi\delta\Phi}
                  \frac{\delta \G_2^{(0)}}{\delta \bar c}
\, .
\label{e11}
\end{eqnarray}
The construction can be iterated. Define for $j>0$
\begin{eqnarray}
\Sigma_j \equiv (-1)^j \frac{1}{j!} \int d^4x \, (\bar c^*)^j
\myfrac{\delta^{(j)} S}{\delta \Phi \dots \delta \Phi} \, .
\label{e12}
\end{eqnarray}
while for $j=0$ we set
\begin{eqnarray}
\Sigma_0 = \G_0^{(0)} \, .
\label{e12_nuovo}
\end{eqnarray}
Then
\begin{eqnarray}
\G_{n}^{(0)} = \sum_{j=0}^{n} \Sigma_j
\label{e13}
\end{eqnarray}
is such that
\begin{eqnarray}
{\cal S}(\G_n^{(0)}) & = &
	 (-1)^n \frac{1}{n!} \int d^4x \,
                   (\bar c^*)^n \myfrac{\delta^{(n+1)} S}
					{\delta \Phi \dots \delta \Phi} s\Phi
		     \nonumber \\
	       & = & (-1)^n \frac{1}{n!} \int d^4x \,
                     (\bar c^*)^n \myfrac{\delta^{(n+1)} S}
				{\delta \Phi \dots \delta \Phi}
                          \frac{\delta \G_n^{(0)}}{\delta \bar c}
                     \, .
\label{e14}
\end{eqnarray}
Thus the breaking of the classical ST identity for the functional
$\G^{(0)}_n$ is of order $n$ in $\bar c^*$.
Therefore the full on-shell classical action, defined by
\begin{eqnarray}
\G^{(0)} = \sum_{j=0}^\infty \Sigma_j \, ,
\label{e_on_shell_class_act}
\end{eqnarray}
fulfills the classical ST identity
\begin{eqnarray}
{\cal S}(\G^{(0)}) = 0 \, .
\label{e_cl_on_shell}
\end{eqnarray}
In particular, if $S[\Phi,\beta]$ is a
polynomial of degree $M$ in $\Phi$,
 the classical action $\G_M^{(0)}$ defined according to eq.~(\ref{e13})
 fulfills the ST identity
\begin{eqnarray}\label{e15}
 {\cal S} (\G_M^{(0)}) = 0 \, .
\end{eqnarray}
Notice that already at the classical level there are mixed couplings
$\bar c^*$-$\beta$, due to the dependence of $\der{S}{\Phi}$ on $\beta$.
This does not prevent the identification of $\beta$ with the source
coupled to the operator ${\cal O}(\Phi)$, since
\begin{eqnarray}
\left . \myfrac{\delta \G_M^{(0)}}{\delta \beta} \right |_{\beta=\bar c^*=0}
=
{\cal O}(\Phi) \, .
\label{e15bis}
\end{eqnarray}
%

\subsection{Quantization}\label{SETon:shell:quant}

We study the quantization of the on-shell ET STI.
We will show that it can be fulfilled by a suitable choice of local
finite counter-terms order by order in the loop
expansion~\cite{Picariello:2001ri}.

The proof is a recursive one.
 Let us assume that
the ST identity has been restored up to order $n-1$, so that
\begin{eqnarray}
{\cal S}(\G)^{(j)} =0 \, , ~~~~~~~~~ j=0,1,\dots,n-1 \, .
\label{en1}
\end{eqnarray}
The $n$-th order ST breaking term $\Delta^{(n)}$, defined by
\begin{eqnarray}
\Delta^{(n)} \equiv {\cal S}(\G)^{(n)} \, ,
\label{en2}
\end{eqnarray}
fulfills the following Wess-Zumino
consistency condition:
\begin{eqnarray}\label{en3}
{\cal S}_0 \Delta^{(n)} = 0 \, ,
\end{eqnarray}
where ${\cal S}_0$ is the classical linearized ST operator given by
\begin{eqnarray} \label{e15ter}
{\cal S}_0(X) \equiv  \int d^4x \,
 \left ( c \myfrac{\delta X}{\delta\varphi} + \theta \myfrac{\delta
X}{\delta\rho}
 \right ) + (\G^{(0)},X) + (X,\G^{(0)}) \, .
\end{eqnarray}
Moreover thanks to the QCAP $\Delta^{(n)}$  is a local integrated formal
power
 series with FP-charge $+1$. Since the only fields and
 external sources with FP-charge $+1$ are $c$ and $\theta$,
 we get that
\begin{eqnarray}
\left . \Delta^{(n)} \right |_{c=\theta=0} = 0 \, .
\label{en4}
\end{eqnarray}
Notice that both $c$ and $\theta$ enter into doublets under the
 classical linearized ST operator ${\cal S}_0$.
However in the present case $(\varphi,c)$ and $(\rho,\theta)$ are
 coupled doublets, according to the terminology of~\cite{doppietti},
 since their counting operator does not commute with ${\cal S}_0$.
Therefore we must resort to the results
 discussed in~\cite{doppietti} and summarized in
 appendix~{\bf\ref{app:doublets}}
 to carry out the recursive cohomological analysis of the ST breaking
 term $\Delta^{(n)}$.

By using these results we get that eq.~(\ref{en4}) combined with
eq.~(\ref{en3})
 guarantees that $\Delta^{(n)}$ belongs to the trivial cohomology
 class of the operator ${\cal S}_0$, namely there exists a local
 integrated formal power series $\Xi^{(n)}$ such that
\begin{eqnarray}
{\cal S}_0(-\Xi^{(n)}) = \Delta^{(n)} \, .
\label{en5}
\end{eqnarray}
By adding the local (in the sense of formal power series)
finite counterterms $\Xi^{(n)}$ to the $n$-th order effective
action we end up with a new symmetric quantum effective action fulfilling
the ST identity up to order $n$.

We conclude that it is possible to define a full quantum effective action
$\G$
fulfilling the ST identity
\begin{eqnarray}
{\cal S}(\G) = 0 \, ,
\label{en6}
\end{eqnarray}
to all orders in the loop expansion.
We remark that $\G$ also depends on the external sources
$\beta$ coupled to local composite operators ${\cal O}(\Phi)$ which are
functions only of $\Phi$. Notice that, unlike in the standard BRST
treatments of gauge theories, the sources $\beta$
are not coupled here to BRST invariant operators.
All the same we will be able to prove the $\rho$-independence of the vertex
functions
\begin{eqnarray}
\myfrac{\delta^{(n)} \GT_{on}}{\delta \beta_{i_1}(x_1) \dots \delta
\beta_{i_n}(x_n)}
\label{e16}
\end{eqnarray}
as a consequence of the ET STI in eq.~(\ref{en6}), once all
the external sources are switched off and the conditions
$\myfrac{\delta \GT_{on}}{\delta c}$=0, $\myfrac{\delta \GT_{on}}{\delta \bar c}$=0 and
$\myfrac{\delta \GT_{on}}{\delta \varphi}$=0 are imposed.

\section{Consequences for the Green functions of the model}
\label{sec:rho:indep}

We now have to investigate the consequences of the ET STI on the
Green functions of the model.
We have identified for each case of the ET a selected set of
Green functions:
\\
{\bf a)} for the WET those of the BRST invariant operators, including
operators dependent only
 on $\Phi(\varphi;\rho)$ (since in this formalism $s\Phi=0$);
\\
{\bf b)} for the off-shell SET those of the BRST invariant operators,
including operators
 dependent only on $\chi$ (since in this formalism $s\chi=0$);
\\
{\bf c)} for the on-shell SET those of the operators dependent only on
 $\Phi(\varphi;\rho)$. 
Notice that in this case these are not BRST
 invariant.
The external source coupled to this operator must be included in the 
 Batalin-Vilkovisky formalism.

In each case the relevant ET STI imply that the above mentioned Green
 functions of local (in the sense of formal power series)
 operators do not depend on $\rho$.

In some cases this set of Green functions can be identified
 with the physics of the model. However this is not always possible.
For instance it can be shown, in~\cite{workinprogress}, that the
 ET provides a somewhat natural quantization prescription for those
 gauge theories which are anomalous in the ordinary BRST quantization.
Although the ET STI can be restored in these models to all orders
 in the loop expansion (assuming that the QCAP holds), their content
 is much weaker than that of ordinary STI based on gauge symmetry.
In particular the former, unlike the latter, say
 nothing about the unitarity of the theory, which is indeed violated
 at the quantum level.
 Therefore we cannot use ET STI to study physical observables.

For this reason we limit ourselves to the discussion of the
 $\rho$-independence of the relevant Green functions discussed
 above, with no claim that they identify the physics of the model.

\par

We work out in detail the proof of the independence of $\rho$ for the
 Green functions of operators dependent only on $\Phi$ for the
 on-shell SET case.
The same pattern applies to the other two cases, where the proof boils
 down to a straightforward paraphrase of standard
 arguments~\cite{Piguet:1995er},
 previously used to discuss the dependence of the BRST invariant
 operators on the gauge parameters in ordinary gauge theories.
Instead in the on-shell case we are dealing with operators which are
 not BRST invariant.
However it turns out that a similar proof can be applied also in this case.

We start from the ST identity  for the connected generating functional
 $\tilde W_{on}$
\begin{eqnarray}\label{i23}
{\cal S}(\tilde W_{on}) &\equiv& \int d^4x \left (
- K(x) \der{\tilde W_{on}}{\bar \eta(x)}
- \eta(x) \der{\tilde W_{on}}{\bar c^*(x)}
+ \theta_i(x)\der{\tilde W_{on}}{\rho_i(x)} \right ) =
0
 \, .
\end{eqnarray}
We differentiate the above equation with respect to $\theta_k(y)$ and get
\begin{eqnarray}
\der{\tilde W_{on}}{\rho_k(y)} = \int d^4x \, \left (
  K(x) \der{^2 \tilde W_{on}}{\theta_k(y) \delta \bar\eta(x)}
+ \eta(x) \der{^2 \tilde W_{on}}{\theta_k(y) \delta \bar c^*(x)}
+ \theta_i(x) \der{^2 \tilde W_{on}}{\theta_k(y) \delta \rho_i(x)}
  \right ) \, .
\label{i26}
\end{eqnarray}
Differentiation of eq.~(\ref{i26}) with respect to the
 sources $\beta_{l_1}(z_1), \dots ,\beta_{l_n}(z_n)$ gives,
 once we go on shell and set $K=\theta=\beta=\eta=\bar\eta=\bar c^*=0$:
\begin{eqnarray*}
\left . \der{^{(n+1)} \tilde W_{on}}
            {\beta_{l_1}(z_1) \dots \delta \beta_{l_n}(z_n)
\delta \rho_k(y)}
\right |_{K=\theta=\beta=\eta=\bar\eta=\bar c^*=0} =0 \, .
\end{eqnarray*}
Therefore the Green functions of operators depending only on
 $\Phi(\varphi;\rho)$ are  independent of $\rho_k$.
This is the content of the on-shell SET.
\section{Example: an application of the ET in the Abelian gauge theory}\label{sec:example}
We want to illustrate our discussion by giving an example
obtained from the Abelian gauge theory.
In this section we show that there exists a very simple field redefinition
 connecting Proca's gauge, which looks unitary but is not renormalizable by
 power-counting, to the Lorentz-covariant gauge, which is renormalizable by
 power-counting.
This procedure based on ET might give a meaning to the massive QED in
 Proca's gauge, which is non-renormalizable in the conventional
 sense.

%
\def\lgamma{\hbox{\large$\gamma$}}
\newcommand{\slsh}[1]{{#1} \!\!\!\!\slash }
\def\slshD{\slsh{\cal D}}
\def\slshA{\slsh{A}}
\def\slshpart{\slsh{\partial}}
%

\subsection{The starting point: the Lorentz-covariant
gauge}
We start with the Lagrangian including matter fields with a gauged $U(1)$
 symmetry, where the gauge boson is massive and is quantized in the
 Lorentz-covariant gauge, parameterized by $\alpha$:
\begin{eqnarray}\label{eq:lag}
S_\alpha =  \int d^4x\, {\cal L}_{\alpha}&=&
\int d^4x\,\left(
 - \frac{1}{4} F_{\mu\nu}F^{\mu\nu}
 + \frac{m^2}{2} A^2
 - \frac{\alpha}{2} (\partial A)^2
 + {\cal L}_{\psi}(A_\mu,\psi,\bar\psi) \right)\,.
\end{eqnarray}
\begin{eqnarray}\label{eq:psi}
{\cal L}_{\psi} = \bar\psi \slshD \psi - m \bar\psi\psi
\end{eqnarray}
is the term containing the fermion fields, which are
minimally coupled to the gauge boson
 (i.e. $\slshD = \slshpart + i e \slshA$).

\subsection{From the Lorentz-covariant
gauge to Proca's gauge: a field redefinition}
We can use the Equivalence Theorem at the classical level to connect a
 theory with mass $m$ and gauge parameter $\alpha$ to a theory with
 the same mass $m$ and gauge parameter $\alpha=0$.
That is, we are dealing with a field transformation connecting
 the Lorentz-covariant gauge to Proca's gauge.

The gauge provided by the action in eq.~(\ref{eq:lag}) can also
be obtained by using the auxiliary field $\varphi$:
\begin{eqnarray}\label{eq:lagphi}
S &=& \int d^4x\, {\cal L}(x)
\nonumber\\
&=& \int d^4x\,\left(
 - \frac{1}{4} F_{\mu\nu}F^{\mu\nu}
 + \frac{m^2}{2} A^2
 + m\sqrt{\alpha}\varphi\partial A
 + \frac{m^2}{2} \varphi^2
 + {\cal L}_{\psi}(A_\mu,\psi,\bar\psi)\right)\, .
\end{eqnarray}
The field $\varphi$ does not enter in the interaction vertices.
Thus we can impose the equation of motion for $\varphi$:
\begin{eqnarray}
\der{S}{\varphi} = m \sqrt{\alpha} \partial A + m^2 \varphi = 0
\label{eq:e33}
\end{eqnarray}
so that
\begin{eqnarray}
\varphi = - \frac{\sqrt{\alpha}}{m} \partial A \, .
\label{eq:e34}
\end{eqnarray}
Both equations (\ref{eq:e33}) and (\ref{eq:e34}), being linear
in the fields, are valid at the quantum level.

If we perform the following field redefinition in eq.~(\ref{eq:lagphi})
\begin{eqnarray}\label{eq:transformationAll}
&& A_\nu \rightarrow
A'_\nu \equiv A_\nu + \frac{\sqrt{\alpha}}{m} \partial_\nu \varphi \, ,
~~~~
\varphi \rightarrow \varphi \, , \nonumber \\
&&
\psi \rightarrow
\psi' \equiv \exp \left ( i e\frac{\sqrt{\alpha}}{m} \varphi \right ) \psi
\,,
~~
\bar\psi \rightarrow
\bar\psi'\equiv\exp\left(-i e\frac{\sqrt{\alpha}}{m} \varphi \right )
\bar\psi \,, \nonumber \\
\label{eq:e31}
\end{eqnarray}
 we obtain the transformed action
\begin{eqnarray}
S_{Proca} &=& \int d^4x\,{\cal L}_{Proca}(x)
\nonumber\\
 &=&\int d^4x\,\left(
 -\frac{1}{4}F_{\mu\nu}F^{\mu\nu}
  + \frac{{m}^2}{2} {A}^2_\mu
  + {\cal L}_{\rm int}(A_\mu,\psi,\bar\psi)
  + \alpha \partial_\nu \varphi\partial^\nu\varphi
  + \frac{m^2}{2}\varphi^2
 \right) \, ,
\label{eq:e32}
\end{eqnarray}
which is the QED action in Proca's gauge with the addition of the
 free scalar field $\varphi$.

\subsection{Proca's gauge}

Notice that in a massive Abelian gauge theory with zero gauge-fixing
parameter the propagator given by
\begin{equation}
D^{\mu\nu}= 
 -i \frac{ g^{\mu\nu}-\cfrac{p^\mu p^\nu}{m^2}
         }{p^2 - m^2 +i\epsilon}\,,
\label{eq:e29}
\end{equation}
 behaves asymptotically as  a constant for large $p^2$ and
 propagates only transverse polarizations.
In the Lorentz-covariant gauge 
\begin{eqnarray}
D^{\mu\nu}_{\alpha}(p) =
-i \frac{g^{\mu\nu} - \frac{p^\mu p^\nu}{m^2}}{p^2-m^2+i\epsilon}
-i\frac{\frac{p^\mu p^\nu}{m^2}}{p^2-\frac{m^2}{\alpha}+i\epsilon}
\,
\label{e7}
\end{eqnarray}
 behaves as $1/p^2$ at large momenta and propagates both spin 1 and
 (unphysical) spin zero modes.
As a consequence, the latter theory can be considered renormalizable
 by power-counting but not manifestly unitary, while the former is not
 renormalizable by power-counting but is manifestly  unitary.

Therefore the field redefinitions in eq.~(\ref{eq:transformationAll})
 connect a unitary non-power-counting renormalizable theory to a
 power-counting renormalizable one.

\subsection{The Equivalence Theorem and Proca's gauge}

We now want to understand what happens at quantum level, i.e.
 we apply the Equivalence Theorem to the transformations in
 eq.~(\ref{eq:transformationAll}).
We introduce a parameter $\rho$ controlling the field redefinitions
 in eq.~(\ref{eq:transformationAll}) and rewrite them as
\begin{eqnarray}
A_\nu &\rightarrow&
 A'_\nu \equiv A_\nu +\rho \frac{\sqrt{\alpha}}{m} \partial_\nu \varphi \, ,
\nonumber\\
 \psi &\rightarrow&
 \psi' \equiv \exp \left ( ie\rho\frac{\sqrt{\alpha}}{m}\varphi\right ) \psi
\,,
\nonumber\\
\bar\psi &\rightarrow&
\bar\psi'\equiv\exp\left(-ie\rho\frac{\sqrt{\alpha}}{m}\varphi\right)
\bar\psi \,.
\label{eq:e8}
\end{eqnarray}
Since the Jacobian of the above field redefinition is one  and the
 transformation is locally invertible, we can apply the WET.

The ET BRST transformation of the parameter $\rho$ is provided by
\begin{eqnarray}
s \rho =\theta \, , ~~~~  s \theta =0 \, ,
\label{e9}
\end{eqnarray}
where $\theta$ is an anti-commuting parameter. The FP-charge of
$\rho,\theta$
is defined by setting $FP(\rho)=0$, $FP(\theta)=1$.

According to the formulation developed in sect.~{\bf\ref{sec:WET}}
for the WET we impose that 
$A'_\mu$, $\psi'$, $\bar\psi'$ are BRST invariant
(see eq.~(\ref{eq:sPhi})). In this way we
 obtain the BRST transformations of $A_\mu,\psi,\bar\psi$
 (denoted by $sA_\mu=\omega_\mu$, $s\psi= \omega$, $s\bar\psi=
 \bar\omega$), in agreement with eq.~(\ref{eq:BRSTWET}):
\begin{eqnarray}
s A'_\mu &=&
 \omega_\mu + \theta \frac{\sqrt{\alpha}}{m} \partial_\mu \varphi = 0 \, ,
 \nonumber \\
s \psi' &=&
  i e\theta\frac{\sqrt{\alpha}}{m}\varphi
	\exp \left ( i e \rho\frac{\sqrt{\alpha}}{m}\varphi\right )\psi
  + \exp \left ( i e \rho\frac{\sqrt{\alpha}}{m}\varphi\right )\omega = 0\,
,
 \nonumber \\
s \bar\psi' &=&
  -i \bar\psi\exp \left (-i e \rho\frac{\sqrt{\alpha}}{m}\varphi\right )
	e \theta\frac{\sqrt{\alpha}}{m}\varphi
  + \bar\omega\exp \left (-i e \rho\frac{\sqrt{\alpha}}{m}\varphi\right ) =
0\, .
\label{e10}
\end{eqnarray}

The solution for $\omega_\mu$ is
\begin{eqnarray}
\omega_\mu = -\theta \frac{\sqrt{\alpha}}{m} \partial_\mu \varphi
\label{et::gaugetrsf}
\end{eqnarray}
which is linear in the fields and does not require the introduction of any
 external source.

Notice that the solution for $\omega$ and $\bar\omega$
\begin{eqnarray}
\omega =
  -i e \theta\frac{\sqrt{\alpha}}{m}\varphi\psi
&\quad\mbox{and}\quad&
\bar\omega=
  i e \bar\psi\theta\frac{\sqrt{\alpha}}{m}\varphi\,
\label{et::matttrsf}
\end{eqnarray}
 is quadratic in the fields.  Since 
 the field
 $\varphi$ is not interacting, according to eq.~(\ref{eq:e33}),
 the products of fields in eq.(\ref{et::matttrsf}) are well-defined 
 and do not require the introduction of the antifields for $\psi,\bar \psi$.
 Thus we can write the ET STI as:
\begin{eqnarray}\label{et::procasti}
&&
\!\!\!\!\!\!\!\!\!\!\!\!
{\cal S}(\G)=\int d^4x\,\left(
 -\theta \frac{\sqrt{\alpha}}{m} \partial_\mu \varphi
	\der{\G}{A_\mu}
 -i e \theta\frac{\sqrt{\alpha}}{m}\varphi\psi
	\der{\G}{\psi}
 +i e \bar\psi\theta\frac{\sqrt{\alpha}}{m}\varphi\,
	\der{\G}{\bar\psi}\right)
 +\theta\der{\G}{\rho}=0
\,.
\end{eqnarray}
We wish to comment on the above equation.
Eq.(\ref{et::procasti}) looks like a Ward identity taking into account
 the gauge transformation of $A_\mu$, $\psi$ and $\bar\psi$.
The last term of ET STI allows to control the dependence of all Green
functions
 on the gauge parameter $\rho$.
Eq.(\ref{et::procasti}) guarantees that the Green functions of
gauge-invariant operators are independent of $\rho$ and thus 
we can look at Proca's gauge as the limit of $\rho$ going to zero.

\section{Concluding remarks}\label{sec:sez6}

In this note we discussed the Equivalence Theorem (ET) in the BRST
formalism.
\par
Some discussion has been devoted to the kind of transformation of the
fields:
if the transformation has a local inverse (at least perturbatively)
then some short-cut in the proof of the ET at quantum level can be used,
but also if the inverse is non-local we proposed some alternative which
allow
 the loop expansion both in the original and in the transformed theory.
\par
We studied the quantum deformation to the associated ST identity  and
 showed that suitably defined sets of Green functions do not depend
 on the choice of the transformation of the fields.
The computation of the cohomology for the linearized ST identity has been
 performed by purely algebraic methods and no power-counting arguments have
 been used.
The results of the paper are based on the conjecture of 
 the Quasi Classical Action Principle. 
\par
As an example we studied 
the massive QED in
 Proca's gauge by relating it to a renormalizable 
gauge theory in the Lorentz-covariant gauge via the
Equivalence Theorem.

\section*{Acknowledgments}
Two of us would like to thank Professor R.~Stora for useful discussions.
The MIUR, Mi\-ni\-ste\-ro dell'Istruzio\-ne dell'Universit\`a e della Ricerca,
Italy,
 and the INFN (sezione di Milano) are acknowledged for the financial
support.

\appendix

\section{The Quasi Classical Action Principle}\label{app:QCAP}

In the framework of Algebraic Renormalization
an essential r\^ole is played by a set of locality properties
of the renormalized Green functions encoded in the
so-called Quantum Action Principle (QAP).
The relevance of the QAP in discussing many aspects
of the renormalized theory by purely algebraic and
regularization-independent methods has emerged since
the first pioneering papers on the BRST quantization of gauge
theories.
It is now believed that the QAP expresses the fundamental locality
requirements
a regularization scheme should possess in order to give rise to a sensible
(local) power-counting renormalizable quantum field theory.
These locality requirements are common to all regularization schemes
and do not depend on the normalization conditions chosen.

Despite this kind of universality, the proof of the QAP has to be given
independently for each regularization procedure.
In the seventies the QAP has been proven
for a variety of regularization schemes,
 including BPHZ and analytical and dimensional
regularization.
We wish to recall here the quite noticeable exception of the
Epstein-Glazer formalism, for which no satisfactory proof of the QAP
has been worked out up to now.

It is customary to summarize the content of the QAP by
characterizing the behavior of the renormalized quantum effective action
$\G$  under infinitesimal variations of the fields and the parameters of the
model. The standard formulation of the QAP (as given
for instance in~\cite{Piguet:1995er})
is the following.

\props{QAP}{
Let $\G$ be the vertex functional corresponding to a
(power-counting renormalizable) theory in a $D$-dimensional
space-time with a classical action given by
\begin{eqnarray}
\G^{(0)}=\int\,d^Dx\, {\cal L}(\varphi_a,\beta_i,\lambda)
\label{eq:classact}
\end{eqnarray}
where $\varphi_a$ are  the quantum fields,
 $\beta_i$ the external sources coupled to field polynomials
 $Q^i$
and $\lambda$ stands for the parameters of the model
 (masses, coupling constants, renormalization points).
Let the inverse of the quadratic part of the action be the standard
 Feynman propagators.\\
Given the local operator
\begin{eqnarray}\label{eq:QAP0}
\mathfrak{s}({\G}) \equiv
\alpha_a \der{\G}{\varphi_a(x)}
+ \alpha_{ab}\varphi_b(x) \der{\G}{\varphi_a(x)}
+ \alpha_{ia}\der{\G}{\beta_i(x)}\der{\G}{\varphi_a(x)}
+ \alpha\der{\G}{\lambda},
\end{eqnarray}

where
$\alpha_a, \alpha_{ab}, \alpha_{ia}$ and $\alpha$ are constants,
then the Quantum Action Principle can be stated in the
following way
\begin{eqnarray}\label{eq:QAP1}
\mathfrak{s}({\G})= \QAPX{ } = \Delta^{(n)}(x)
+ O(\hbar^{n+1}).
\end{eqnarray}

$\QAPX{ }$ denotes the insertion of a local operator.
Moreover the lowest non-vanishing order coefficient $\Delta^{(n)}(x)$
of $\QAPX{ }$ is a local
polynomial in the fields  and external sources and their derivatives
with bounded dimension.

}

At the integrated level (Slavnov-Taylor-like identities) the QAP
reads
\begin{eqnarray}
{\cal S}(\G) \equiv \int d^4x \, \mathfrak{s}(\G) = \int d^4x \,
\QAPX{ } \, .
\label{eq:QAP2}
\end{eqnarray}
The first non-vanishing order of the ST-like breaking terms is
given by
\begin{eqnarray}
\Delta^{(n)} \equiv {\cal S}(\G)^{(n)} = \int d^4x \, \Delta^{(n)}(x) \, .
\label{eq:intSTI}
\end{eqnarray}
As a consequence
of Proposition~\ref{prop:QAP},
$\Delta^{(n)}$
 is an integrated local polynomial in the fields and the external
sources and their derivatives with bounded dimension.

The ultraviolet (UV) dimension $d_\QCAP{ }$ of $\QCAP{ }^{(n)}$ can be
predicted from the UV dimensions
$d_a$ of the fields $\varphi_a$ and the UV dimensions $d_{Q^i}$ of
the field polynomials $Q^i$~\cite{Piguet:1995er}.
We do not dwell on this problem here
since the only information we need for the present discussion is the
fact that  $d_\QCAP{ }$ is bounded.

The QAP tells us that in a power-counting renormalizable theory
the ST-like identity in eq.~(\ref{eq:QAP2}) can be broken at quantum level
 only by the insertion of an integrated
 local composite operator of bounded dimension.
This is an all-order statement holding true regardless the normalization
conditions chosen.
At the lowest non-vanishing order the insertion $\QAPX{ }$ reduces to a
 local polynomial in the fields and external sources and their derivatives
 with bounded dimensions.
This property is a consequence of the topological nature of the
 $\hbar$-expansion as a loop expansion.
That is, if a local insertion in the vertex functional
were zero up to the order $n-1$, at the $n$-th order it must reduce
from a diagrammatic point of view to a  set of points.
By power-counting this set is  finite and hence
it corresponds to a local
polynomial in the fields and the external sources and their
derivatives.

The extension of the QAP beyond the power-counting renormalizable case
is yet an open issue in the theory of renormalization. For a
non-power-counting
(but still polynomial) interaction Lagrangian it is known in
analytical renormalization that the breaking of the ST-like
identities is given by the insertion of a (possibly infinite) sum of local
operators.
Quite restrictive assumptions on the form of the propagators are
needed in the proofs of both the QAP and extensions thereof.
In all cases the key ingredient of these diagrammatic analyses is
Zimmerman's forest formula, allowing to treat in a systematic way
the rather formidable combinatorial intricacies of the renormalization
procedure.

A complementary approach has been recently formulated in
\cite{Stora2000} by Stora. A statement weaker than the QAP
has been conjectured to hold true in very general situations,
as a consequence of first principles of (axiomatic)
local quantum field theory.
Unlike in the QAP no reference is made to the all-orders
characterization of ST-like breaking terms. It is only stated
that the first non-vanishing order in the loop expansion is a local
functional. Here locality should be generally understood
in the sense of local formal power series, since the loss of power-counting
forces to drop bounds on the dimensions.

The starting point is the existence of
 an off-shell $S$-matrix fulfilling
 causal factorization. We also assume that the free fields can
be derived from a
 Lagrangian, which in turn implies (in the presence of a mass gap)
the validity of the LSZ reduction formulae.
The latter allow to make the connection with the usual
Green functions by defining their
 generating functional $Z[J,g]$ to be equal to the
vacuum expectation value of the off-shell $S$-matrix.
Here $J$ is a collective notation for the external sources
linearly coupled
to the quantized fields $\{\varphi_a\}$ and $g$ stands for all the external
sources $\beta$ coupled to composite operators and for all the coupling
 constants $\lambda$
 (promoted to test functions in the spirit of the Epstein-Glazer approach).

Stora's conjecture can the be summarized as follows \cite{Stora2000}:

\props{QCAPStora}{
Let
$${\cal D}(\epsilon)=\int d^4\,x\, \epsilon(x){\cal D}_x[J,g]$$
where ${\cal D}_x[J,g]$ is a local functional first order linear
differential
 operator:
$$
{\cal D}_x[J,g] ={\cal G}(J,g)(x)\der{}{g(x)} +  K (J,g)(x)\der{}{J(x)}
$$
 with ${\cal G}$ and $ K $ are local in $g$ and $J$, and $\epsilon(x)$
 is a test function.
\\
Let
$$
W[J,g] = \sum_n \hbar^n W^{(n)}[J,g]
$$
be the connected generating functional expanded in powers of the
loop-counting
parameter $\hbar$.
\\
Assume that
$$
{\cal D}(\epsilon)W^{(i)}[J,g] = 0 \text{ for } i<n
$$
Then
$$
\left.
{\cal D}(\epsilon)W^{(n)}[J,g] = \Delta^{(n)}[J,g,\varphi](\epsilon)
\right|_{\frac{\delta W^{(0)}[J,g]}{\delta J(x)}=\varphi(x)}
$$
for some $\Delta^{(n)}$ local in $(J,g,\varphi)$.
}

In this approach the diagrammatic analysis is avoided and no mention is made
to the detailed structure of the interaction Lagrangian of ordinary
perturbative quantum field theory. Moreover in this formulation
there is no reference to
 the concrete procedure allowing to construct the $n$-th order in the loop
expansion,
 starting from lower orders. In particular we do not rest on any
 specific regularization procedure.
In this sense the conjecture captures the universal features of the locality
properties one should demand for any physically sensible
renormalization. This in turn renders the content of the conjecture
the natural expectation to be required for all possible
extensions of ordinary quantum field theory to those situations
where power-counting renormalizability might not hold.

Up to now no proof of Proposition~\ref{prop:QCAPStora} is known.
In the present paper we assume its validity and make extensive use
 of the following consequence of Proposition~\ref{prop:QCAPStora},
 known as the Quasi-Classical Action Principle:

\props{QCAP}{
Let $\G$ be the vertex functional corresponding to a
 theory 
 with a classical action given by
$$\G^{(0)}=\int\,d^4x\, {\cal L}(\varphi_a,\beta_i,\lambda)$$
where $\varphi_a$ are the quantum fields,
 $\beta_i$ the external sources coupled to field polynomials
 $Q^i$, and $\lambda$ stands for the parameters of the model
 (masses, coupling constants, renormalization points).
We assume that the Legendre transform $W[J,\beta,\lambda]$ of
$\G[\varphi,\beta,\lambda]$
\begin{eqnarray}
W[J,\beta,\lambda] = \G[\varphi,\beta,\lambda] + \int d^4x \, J(x)
\varphi(x) \, ,
~~~~ J(x)=-\myfrac{\delta \G}{\delta \varphi(x)} \, , ~~~~
\myfrac{\delta \G}{\delta \beta(x)} = \myfrac{\delta W}{\delta \beta(x)}
\, ,
\label{eq:legtrans}
\end{eqnarray}
fulfills Proposition~\ref{prop:QCAPStora}.
Notice that $\G^{(0)}$ is not required to be power-counting renormalizable.
Let us also define
\\
\begin{eqnarray}
\mathfrak{s}({\G}) &=&
\alpha_a \der{\G}{\varphi_a(x)}
+ \alpha_{ab}\varphi_b(x) \der{\G}{\varphi_a(x)}
+ \alpha_{ia}\der{\G}{\beta_i(x)}\der{\G}{\varphi_a(x)}
+ \alpha\der{\G}{\lambda}\, .
\end{eqnarray}
Then {\rm (\bf Quasi Classical Action Principle\rm)}
the first non-vanishing order in the loop expansion, say $n$,
of $\mathfrak{s}(\G)$:
\begin{eqnarray}
\Delta^{(n)} \equiv \mathfrak{s}(\G)^{(n)} \, , ~~~~~
\mathfrak{s}(\G)^{(j)}=0 \ \text{for } j=0,1,\dots,n-1
\label{eq:qcapbrkg}
\end{eqnarray}
is a local formal power series in the fields and external sources
and their derivatives.
}

In the power-counting renormalizable case
bounds on the dimensions can be given truncating the formal power series
predicted by the QCAP to a local polynomial. Thus
the QCAP reduces in this case to the
part of the QAP stating that the lowest non-vanishing order
$\Delta^{(n)}(x)$ of $\Delta(x)\cdot \G$ in eq.~(\ref{eq:QAP1}) is a
polynomial.
This justifies the name of Quasi-Classical Action Principle.

For most practical purposes the QCAP (or, for power-counting renormalizable
theories, the part of the QAP relevant to $\Delta^{(n)}(x)$) is what is
really needed in order to carry out the program of Algebraic
Renormalization.
In particular, this is enough to discuss the restoration of
anomaly-free ST-like identities order by order in the loop expansion.
This point is illustrated in the present paper on the example
of the quantization of the Equivalence Theorem ST identities.

\section{Cohomology of coupled doublets}\label{app:doublets}

The computation of the cohomology of an arbitray nilpotent
differential $s$ (including BRST differentials and classical linearized
 Slavnov-Taylor operators) can be
simplified thanks to the elimination of the fields
and external sources entering as doublets under $s$.

We make here the distinction between (decoupled) doublets,
in the ordinary sense used in
literature, and coupled doublets.

\definitions{A pair of variables ($\rho$,$\theta$) is called a
(decoupled) doublet under the nilpotent differential $s$ if and only if
\begin{eqnarray}
s \rho = \theta\,,\quad s \theta=0
\label{dop1}
\end{eqnarray}
and their counting operator
\begin{eqnarray}
{\cal N} = \sum_k \int d^4x \, \left
(
\rho_k \der{}{\rho_k} +
\theta_k \der{}{\theta_k}
\right )
\end{eqnarray}
commutes with $s$:
\begin{eqnarray}
\left[{\cal N},s\right] = 0 \, .
\label{dop2}
\end{eqnarray}
}
The last condition justifies the name of decoupled doublets.
Usually in the literature
the variables fulfilling the  conditions in eqs.(\ref{dop1}) and
(\ref{dop2}) are simply called ``doublets''.
Here we wish to charcterize them further as ``decoupled'' since we
will drop soon the condition in eq.~(\ref{dop2}).
It is a well-known fact  that
\lemmas{doubletAll}{
If a pair of variables ($\rho$,$\theta$) is a
decoupled doublet under the nilpotent differential $s$
and ${\cal I}[\rho,\theta,\zeta]$ is a $s$-invariant
local integrated formal power series depending
on $\rho,\theta$ and on other variables collectively denoted by $\zeta$,
then its $(\rho,\theta)$-dependent part is $s$-closed:
$${\cal I}[\rho,\theta,\zeta] - {\cal I}[0,0,\zeta] =
s{\cal G}[\rho,\theta,\zeta] \,  $$
for some local integrated formal power series ${\cal G}[\rho,\theta,\zeta]$.
}

Proofs of the above Lemma can be found for instance
in \cite{Piguet:1995er} and in \cite{henneauxreview}.
As a consequence of \lem{doubletAll}, it can be
proven that the cohomology of $s$ in the space
of local integrated formal power series does not depend on $(\rho,\theta)$
\cite{Piguet:1995er,henneauxreview}.

The external sources $(\rho,\theta)$,
introduced in the discussion of the on-shell formulation
of the Equivalence Theorem, fulfill
eq.~(\ref{dop1}), where the relevant nilpotent differential $s$ is now
the classical linearized ST operator (see eq.~(\ref{eq:a:e8}) for the WET
and (\ref{e15ter}) for the SET).
However they do not meet eq.~(\ref{dop2}).
We introduce then the following definition:
\definitions{A pair of variables ($\rho$,$\theta$) is called a
``coupled doublet'' under the nilpotent differential $s$ if
\begin{eqnarray*}
s \rho = \theta\,,\quad s \theta=0
\end{eqnarray*}
and their counting operator
\begin{eqnarray}
{\cal N} = \sum_k \int d^4x \, \left
(
\rho_k \der{}{\rho_k} +
\theta_k \der{}{\theta_k}
\right )
\end{eqnarray}
does not commute with $s$
\begin{eqnarray}
\left[{\cal N},s\right] \neq 0 \, .
\end{eqnarray}
}
It turns out \cite{doppietti} that the cohomology of $s$ in the space
of local integrated formal power series can be completely characterized
in terms of the doublets-independent component of $s$.
However the statement of \lem{doubletAll} is no more true.
A weaker statement, which proves to be sufficient for the discussion
of the ET STI in the on-shell formalism,
can be proven \cite{doppietti}:
\lemmas{doublets}{
Let ${\cal I}[\rho,\theta,\zeta]$ be a local integrated formal power series
closed under the nilpotent
differential $s$:
\begin{eqnarray}
s {\cal I} =0 \, .
\label{dp::de3bis}
\end{eqnarray}
Moreover let us assume that $s{\cal I}[0,0,\zeta]=0$.
Then we have
\begin{eqnarray}
{\cal I}[\rho,\theta,\zeta] - {\cal I}[0,0,\zeta] =
 s {\cal G}[\rho,\theta,\zeta]
\label{dp::B1}
\end{eqnarray}
for some local integrated formal power series ${\cal G}[\rho,\theta,\zeta]$.
}
In particular we remark that eq.~(\ref{dp::de3bis})
is satisfied if
\begin{eqnarray}
{\cal I}[0,0,\zeta]=0 \, .
\label{dop10}
\end{eqnarray}
In  the ET STI the $n$-th order ST breaking term
$\Delta^{(n)}$ plays the r\^ole of ${\cal I}[\rho,\theta,\zeta]$.
$\Delta^{(n)}$ has FP-charge $+1$.
Since
the only fields and external sources with FP-charge $+1$ all enter
as doublets under the linearized classical ST operator, $\Delta^{(n)}$
fulfills eq.~(\ref{dop10}).
Therefore we can apply \lem{doublets}. This in turn allows
to prove that the
on-shell ET STI are non-anomalous, as is explained in
Sect.~\ref{sec:QuantumSTI} and \ref{SETon:shell:quant}.

\end{document}